\documentstyle[12pt]{article}

\def\journal{\topmargin .3in	\oddsidemargin .5in
	\headheight 0pt	\headsep 0pt
	\textwidth 5.625in 
	\textheight 8.25in 
	\marginparwidth 1.5in
	\parindent 2em
	\parskip .5ex plus .1ex		\jot = 1.5ex}
\journal

\catcode`\@=11
\def\marginnote#1{}
\newcount\hour
\newcount\minute
\newtoks\amorpm
\hour=\time\divide\hour by60
\minute=\time{\multiply\hour by60 \global\advance\minute by-\hour}
\edef\standardtime{{\ifnum\hour<12 \global\amorpm={am}%
	\else\global\amorpm={pm}\advance\hour by-12 \fi
	\ifnum\hour=0 \hour=12 \fi
	\number\hour:\ifnum\minute<10 0\fi\number\minute\the\amorpm}}
\edef\militarytime{\number\hour:\ifnum\minute<10 0\fi\number\minute}
\def\draftlabel#1{{\@bsphack\if@filesw {\let\thepage\relax
   \xdef\@gtempa{\write\@auxout{\string
      \newlabel{#1}{{\@currentlabel}{\thepage}}}}}\@gtempa
   \if@nobreak \ifvmode\nobreak\fi\fi\fi\@esphack}
	\gdef\@eqnlabel{#1}}
\def\@eqnlabel{}
\def\@vacuum{}
\def\draftmarginnote#1{\marginpar{\raggedright\scriptsize\tt#1}}
\def\draft{\oddsidemargin -.5truein
	\def\@oddfoot{\sl preliminary draft \hfil
	\rm\thepage\hfil\sl\today\quad\militarytime}
	\let\@evenfoot\@oddfoot	\overfullrule 3pt
	\let\label=\draftlabel
	\let\marginnote=\draftmarginnote
   \def\@eqnnum{(\theequation)\rlap{\kern\marginparsep\tt\@eqnlabel}%
\global\let\@eqnlabel\@vacuum}  }
\def\preprint{\twocolumn\sloppy\flushbottom\parindent 2em
	\leftmargini 2em\leftmarginv .5em\leftmarginvi .5em
	\oddsidemargin -.5in	\evensidemargin -.5in
	\columnsep .4in	\footheight 0pt
	\textwidth 10in	\topmargin  -.4in
	\headheight 12pt \topskip .4in
	\textheight 7.1in \footskip 0pt
	\def\@oddhead{\thepage\hfil\addtocounter{page}{1}\thepage}
	\let\@evenhead\@oddhead	\def\@oddfoot{}	\def\@evenfoot{} }
\def\numberbysection{\@addtoreset{equation}{section}
	\def\theequation{\thesection.\arabic{equation}}}
\def\underline#1{\relax\ifmmode\@@underline#1\else
	$\@@underline{\hbox{#1}}$\relax\fi}
\def\titlepage{\@restonecolfalse\if@twocolumn\@restonecoltrue\onecolumn
     \else \newpage \fi \thispagestyle{empty}\c@page\z@
	\def\thefootnote{\fnsymbol{footnote}} }
\def\endtitlepage{\if@restonecol\twocolumn \else \newpage \fi
	\def\thefootnote{\arabic{footnote}}
	\setcounter{footnote}{0}}  
\catcode`@=12
\relax
\def\figcap{\section*{Figure Captions\markboth
	{FIGURECAPTIONS}{FIGURECAPTIONS}}\list
	{Figure \arabic{enumi}:\hfill}{\settowidth\labelwidth{Figure 999:}
	\leftmargin\labelwidth
	\advance\leftmargin\labelsep\usecounter{enumi}}}
 \relax
\def\tablecap{\section*{Table Captions\markboth
	{TABLECAPTIONS}{TABLECAPTIONS}}\list
	{Table \arabic{enumi}:\hfill}{\settowidth\labelwidth{Table 999:}
	\leftmargin\labelwidth
	\advance\leftmargin\labelsep\usecounter{enumi}}}
 \relax
\def\reflist{\section*{References\markboth
	{REFLIST}{REFLIST}}\list
	{[\arabic{enumi}]\hfill}{\settowidth\labelwidth{[999]}
	\leftmargin\labelwidth
	\advance\leftmargin\labelsep\usecounter{enumi}}}
 \relax
\makeatletter
\newcounter{pubctr}
\def\publist{\@ifnextchar[{\@publist}{\@@publist}}
\def\@publist[#1]{\list
	{[\arabic{pubctr}]\hfill}{\settowidth\labelwidth{[999]}
	\leftmargin\labelwidth
	\advance\leftmargin\labelsep
	\@nmbrlisttrue\def\@listctr{pubctr}
	\setcounter{pubctr}{#1}\addtocounter{pubctr}{-1}}}
\def\@@publist{\list
	{[\arabic{pubctr}]\hfill}{\settowidth\labelwidth{[999]}
	\leftmargin\labelwidth
	\advance\leftmargin\labelsep
	\@nmbrlisttrue\def\@listctr{pubctr}}}
 \relax
\makeatother
\catcode`\@=11
\def\section{\@startsection {section}{1}{0pt}{-3.5ex plus -1ex minus
 -.2ex}{2.3ex plus .2ex}{\raggedright\large\bf}}
\catcode`\@=12
\newskip\humongous \humongous=0pt plus 1000pt minus 1000pt
\def\caja{\mathsurround=0pt}

\newif\ifdtup
\def\panorama{\global\dtuptrue \openup1\jot \caja
	\everycr{\noalign{\ifdtup \global\dtupfalse
	\vskip-\lineskiplimit \vskip\normallineskiplimit
	\else \penalty\interdisplaylinepenalty \fi}}}
\def\eqalignno#1{\panorama \tabskip=\humongous
	\halign to\displaywidth{\hfil$\displaystyle{##}$
	\tabskip=0pt&$\displaystyle{{}##}$\hfil
	\tabskip=\humongous&\llap{$##$}\tabskip=0pt
	\crcr#1\crcr}}

\def\oldreffmt#1{\rlap{[#1]} \hbox to 2\parindent{}}

\def\figfmt#1{\rlap{Figure {#1}} \hbox to 1in{}}

\def\beq{\begin{equation}}
\def\eeq{\end{equation}}

\def\bea{\begin{eqnarray}}

\def\eea{\end{eqnarray}}
\catcode`\@=11
\def\eqnarray{\stepcounter{equation}\let\@currentlabel=\theequation
\global\@eqnswtrue
\global\@eqcnt\z@\tabskip\@centering\let\\=\@eqncr
\gdef\@@fix{}\def\eqno##1{\gdef\@@fix{##1}}%
$$\halign to \displaywidth\bgroup\@eqnsel\hskip\@centering
  $\displaystyle\tabskip\z@{##}$&\global\@eqcnt\@ne
  \hskip 2\arraycolsep \hfil${##}$\hfil
  &\global\@eqcnt\tw@ \hskip 2\arraycolsep $\displaystyle\tabskip\z@{##}$\hfil
   \tabskip\@centering&\llap{##}\tabskip\z@\cr}

\def\@@eqncr{\let\@tempa\relax
    \ifcase\@eqcnt \def\@tempa{& & &}\or \def\@tempa{& &}
      \else \def\@tempa{&}\fi
     \@tempa \if@eqnsw\@eqnnum\stepcounter{equation}\else\@@fix\gdef\@@fix{}\fi
     \global\@eqnswtrue\global\@eqcnt\z@\cr}

\catcode`\@=12
\font\tenbifull=cmmib10 
\font\tenbimed=cmmib10 scaled 800
\font\tenbismall=cmmib10 scaled 666
\relax

\def\thefootnote{\fnsymbol{footnote}}
\def\ref#1{$^{#1)}$}
\begin{document}
\begin{titlepage}
\begin{center}
\today     \hfill    LBL-32856 \\
          \hfill    UCB-PTH-92/35 \\

\vskip .5in

{\large \bf A Candidate For The QCD String Based On Non-Backtracking
Random Walk}
\footnote{This work was supported in part by the Director, Office of
Energy Research, Office of High Energy and Nuclear Physics, Division of
High Energy Physics of the U.S. Department of Energy under Contract
DE-AC03-76SF00098 and in part by the National Science Foundation under
grant PHY90-21139.}

\vskip .5in

Korkut Bardakci \\[.5in]

{\em  Department of Physics\\
      University of California\\
      and\\
      Theoretical Physics Group\\
      Physics Division\\
      Lawrence Berkeley Laboratory\\
      1 Cyclotron Road\\
      Berkeley, California 94720}
\end{center}

\vskip .5in

\begin{abstract}

A new string theory is proposed as a candidate for the  large N limit of QCD.
In this theory, strings are constrained to be non-backtracking; a condition
that is essential for the gauge invariance of the underlying field theory.This
condition is implemented by first placing the theory on a lattice and then
introducing fermionic variables on the world sheet. The naive continuum limit
leads to a generalized Thirring model on the world sheet,and it is suggested
that the application of the renormalization group should drive the coupling
constants to a conformally invariant fixed point.
\end{abstract}
\end{titlepage}
\renewcommand{\thepage}{\roman{page}}
\setcounter{page}{2}
\mbox{ }

\vskip 1in

\begin{center}
{\bf Disclaimer}
\end{center}

\vskip .2in

\begin{scriptsize}
\begin{quotation}
This document was prepared as an account of work sponsored by the United
States Government.  Neither the United States Government nor any agency
thereof, nor The Regents of the University of California, nor any of their
employees, makes any warranty, express or implied, or assumes any legal
liability or responsibility for the accuracy, completeness, or usefulness
of any information, apparatus, product, or process disclosed, or represents
that its use would not infringe privately owned rights.  Reference herein
to any specific commercial products process, or service by its trade name,
trademark, manufacturer, or otherwise, does not necessarily constitute or
imply its endorsement, recommendation, or favoring by the United States
Government or any agency thereof, or The Regents of the University of
California.  The views and opinions of authors expressed herein do not
necessarily state or reflect those of the United States Government or any
agency thereof of The Regents of the University of California and shall
not be used for advertising or product endorsement purposes.
\end{quotation}
\end{scriptsize}

\vskip 2in

\begin{center}
\begin{small}
{\it Lawrence Berkeley Laboratory is an equal opportunity employer.}
\end{small}
\end{center}

\newpage
\renewcommand{\thepage}{\arabic{page}}
\setcounter{page}{1}
\noindent {\bf 1.Introduction}
\vskip 9pt

It has been known for some time that in the large N limit,QCD without matter
fields is equivalent to a free string theory(1).Despite considerable effort(2),
this string theory is yet to be discovered.Recently,there has been renewed
interest in this problem(3),as well as in other approaches to QCD at large N in
general(4).Whatever it is, the QCD string cannot be the same as the standard
Nambu-Polyakov string for at least the following reasons:\\
a)There are no zero mass particles(and of course no tachyon)in the QCD
spectrum.\\
b)The Regge trajectories of QCD should be curved,as opposed to the straight
trajectories of the standard string models.This is true of even two
dimensional QCD coupled to matter(5).

One possible approach to this problem is to reformulate the dynamical equations
of QCD in terms of gauge invariant quantities such as the Wilson loops.These
equations have been derived for both the lattice(6) and the continuum(7)
versions of the theory.Unfortunately,this approach has not thus far led to
a simple string picture.One of the problems with the lattice appraoch is that
the resulting string picture is extremely complicated(8).This is probably due
to the simplicity of the starting point,which is the usual Wilson single
plaquette interaction;it is conceivable that a more complicated interaction,
which is in
the same universality class and which therefore has the same continuum
limit, may give rise to a simpler string.However,since it is not clear a priori
how to determine the interaction that corresponds to a simple string picture
on the lattice,it is tempting to bypass the lattice altogether and work
directly in the continuum with Migdal-Makeenko equations(7).In this case,one
encounters the problem of finding a suitable non-perturbative regulator to
avoid divergences.So far,not much progress has been made along these lines.

In this paper,we bypass the difficult problem of showing directly the
equivalence between QCD and the corresponding string theory;instead,we
focus on the constraints satisfied by Wilson loops.One can think of the
world sheet of the QCD string as being traced by a spacelike Wilson loop
propagating in time,and in the Hamiltonian description of this process,
the physical Hilbert space is generated by the collection of all spacelike
Wilson loops.This forms an overcomplete set of states as a consequence of
 two
kinds of constraints satisfied by the Wilson loops:\\
a)The Mandelstam identities(9).They follow from the fact that N,the dimension
of the color space ,is finite.It is generally assumed that in the limit
 $N\rightarrow\infty$,there are no identities of this type.Since we are
always in the large N limit in this paper,we will not consider them any
further.\\
b)Every backtracking Wilson loop is equivalent to a
non-backtracking one with the backtracking portion removed(see Fig.1).We shall
argue in the next section that this equivalence is crucial for the existence
of an underlying gauge theory;if this condition is not satisfied,Wilson loops
cannot be constructed from a gauge connection.Conversely,if this condition
holds,there always exists a connection valued in some (in general infinite
dimensional) color space.This is the condition we focus on here.

In the standard bosonic string theory,one has an unrestricted sum over all
possible world sheets,and therefore, taking a fixed time slice,all possible
Wilson loops,backtracking as well as non-backtracking,are present.Since the
backtracking loops are equivalent to the corresponding non-backtracking ones
,there is clearly some overcounting.At the minimum,
a string theory based on QCD should avoid this overcounting and sum only on
world sheets whose cross sections are  non-backtracking loops.
In this paper,we will study  precisely such a string theory where the sum over
the world sheets is suitably restricted.Because such a theory always admits a
connection(see (b)above),it corresponds to some kind of gauge theory,and the
 question is,does it have the same interaction as QCD?In this paper,we shall
leave this question open,and study this new string theory as an interesting
theory in its own right.Since the underlying field theory is  gauge invariant,
it is a plausible candidate for the QCD string,hence the title of this paper.
In a future publication(10),we hope to show that  the version of the model
where time is continuous and space is latticized corresponds to a gauge
theory on a space lattice.However,the interaction is not the standard single
plaquette interaction of Wilson, but instead it is a suitably weighed infinite
sum over the traces of all possible Wilson loops. This sum has the standard
 QCD interaction as its naive continuum limit,
and if we believe this naive limit,which we afterall do in the case of the
single plaquette interaction,the equivalence to QCD follows.

The technical problem that will occupy us for most of the rest of this paper
 is how to impose the condition that strings are non-backtracking.Since it is
difficult to handle this
condition directly in the continuum,we first start with a regular space-time
lattice and only at the end take the continuum limit.As a warmup exercise,in
section 2, we
consider the simpler problem of non-backtracking random walk(Brownian motion),
which involves summing over non-bactracking paths on the lattice.This
problem can be solved in a number of different ways;the method which is
easiest to generalize to sums over surfaces  is the standard path integral
approach.In order to implement the non-backtracking condition,we introduce
additional fermionic variables,and we construct the action,making use of the
transfer matrix
method.The path integral must be supplemented by a boundary condition on the
conserved fermion number to eliminate the unwanted backtracking paths.In
section 3,this method is generalized to non-backtracking surfaces,which are
defined as the direct product of two non-backtracking curves,and again a
path integral formulation is obtained.The naive continuum limit of this model
is discussed in section 4;the resulting two dimensional field theory that
lives on the world sheet is a generalized Thirring model with the following
properties:\\
a)It is scale invariant at the classical level.\\
b)It is Lorentz  invariant in the target space.\\
c)It is Lorentz and parity invariant on the two dimensional world sheet as
well.\\
d)There is invariance under both vector and axial phase transformations of
fermions,assuming absence of anomalies.As a result,there exists a chirally
conserved fermion current on the world sheet.Just as in the case of random
walk,suitable
boundary conditions must be
imposed on the conserved fermion current in order to
eliminate the unwanted sectors of the model.

There are several coupling
constants in world sheet field theory that are easily
calculated in the naive continuum limit.Of course,this is not the final
answer;these constants should be allowed to run under the renormalization
group equations and reach a fixed point.We are unable to carry out this
analysis,instead,we conjecture that there is a non-trivial fixed point and
that the theory is conformally invariant at that point.Naturally,in searching
for this fixed point, only those renormalization flows that preserve the
invariances a) to d) should be considered.The main result of this paper is
therefore the following:The original problem  of non-backtracking strings
has been transformed into the
problem of finding the conformal fixed point of a generalized
Thirring model invariant under the symmetries listed above.We feel that this
is a result of interest, although finding such
a fixed point,and solving the resulting model,are difficult problems,and they
will not be attempted here.So far,in problems of this type,solutions
have been found only in very special cases(11), and hopefully,methods
 developed in the study of conformal theories should prove useful in tackling
 this problem in the future.More advanced problems such as the value of the
central charge and the higher order corrections in $1/N$ are also beyond the
scope of this paper.
\vskip 9pt
\noindent {\bf 2.Non-Backtracking Random Walk}
\vskip 9pt

We start by fixing our notation and defining some terminology.Space-time
(target space)is taken to be D dimensional;the  value of D will
not be fixed in this paper.The metric in the target space is flat and
can be either Euclidean or Minkowski.We are interested in both open and
closed strings;we identify the open string with the Wilson line integral
$$
U_{\Gamma}(x,y)=P\{exp\int_{x}^{y}dx'.A(x')\},\eqno(2.1a)
$$
where $\Gamma$ is a path with endpoints x and y,A is the connection valued
in some (infinite dimensional) matrix space,and P denotes path ordering.U
could be thought of as generating a line of flux connecting two external
quarks at x and y along the path $\Gamma$.Closed strings correspond to
$$
W=Tr\{U_{\Gamma}(x,x)\}\eqno(2.1b)
$$
for a closed path $\Gamma$.The equivalence between the U's associated with
the backtracking path $\Gamma$ and the corresponding non-backtracking path
$\Gamma^{'}$ pictured in Fig.1 follows from
$$
U_{\Gamma}(x,y)=U_{\Gamma^{-1}}^{-1}(y,x),\eqno(2.2)
$$
where $\Gamma^{-1}$ is $\Gamma$ traversed in the opposite sense. As a result,
the contributions to U from the segments a and b in Fig.1 cancel. Of course,
(2.2) follows immediately from (2.1a), but the converse is  also true: If eq.
(2.2) holds for arbitrary paths, and in addition, we also require the
semi-group
property
$$
U_{\Gamma^{(1,2)}}=U_{\Gamma^{(1)}}U_{\Gamma^{(2)}}\eqno(2.3)
$$
for two arbitrary paths $\Gamma^{(1)}$ and $\Gamma^{(2)}$ that join at
point $y$
and give $\Gamma^{(12)}$ (see Fig.2),
then U can be written in terms in terms of a
connection A as in (2.1). Therefore,eq's (2.2)and (2.3) guarantee the existence
of an underlying gauge theory. In a random walk(or a string) problem,one has to
deal with the paths directly, instead of the U's attached to them,and so
it is convenient to define an abstract group generated by path multiplication
(12). The group axioms are
$$
\Gamma(x,y)\Gamma^{-1}(y,x)=1\eqno(2.4a)
$$
and
$$
\Gamma^{(1,2)}(x,z)=\Gamma^{(1)}(x,y)\Gamma^{(2)}(y,z),\eqno(2.4b)
$$
where, again, multiplication of two paths consists of joining them beginning
to end to form a new path(Fig.2). The U's can then be thought of as
representations of the abstract group relations given by eqn's (2.4).It is easy
see that, on the lattice,(2.4b) is trivially satisfied if we build up
the path(string)link by link; on the other, the cancellation condition (2.4a)
 is not automatic and remains to be implemented.

Before turning to the problem of non-bactracking random walk,as a preliminary
exercise,we will briefly consider the simpler and extensively studied problem
of random walk without constraints(13). This problem will be solved first by
 the
operator and then by the path integral method. The reason for going through
 this simple exercise is that the methods used
 generalize easily to the real problem. We start with a
regular lattice with spacing d in D dimensions,and label the lattice sites by
the vector ${\bf x}$,whose components are given by $x_{a}$. Boldface letters
 will always denote vectors in the target space. Let the orthonormal
set of states $|{\bf x}>$ be  discrete eigenstates of ${\bf x}$ with the
corresponding eigenvalue,
and let $p_{a}=-i\frac{\partial}{\partial x_{a}}$ be the components of the
conjugate momentum ${\bf p}$. The operator $exp(idjp_{a})$ shifts by one link
 in the positive "a" direction for $j=+1$ and by one link in the negative "a"
direction for $j=-1$. We now consider a sum over all paths starting at
 ${\bf x}$
and ending at ${\bf x'}$, each weighed by a Boltzmann factor $exp(-\alpha l)$,
where l=nd is the length of the path. It is given by the following operator
expression:
$$
G({\bf x},{\bf x'})=<{\bf x'}|(1-2e^{-\alpha d}\sum_{a}cos(dp_{a}))^{-1}
|{\bf x}>.\eqno(2.5)
$$
We are also interested in the continuum limit $d\rightarrow 0$. This limit is
taken by replacing the cosine by
$$
\sum_{a}cos(dp_{a})\approx D-\frac{1}{2}d^{2}\sum_{a}p^{2}_{a}\eqno(2.6a)
$$
and scaling G by
$$
\bar{G}=d^{2-D}e^{-\alpha d}G,\eqno(2.6b)
$$
while keeping $\kappa^{2}=(e^{\alpha d}-2D)/d^{2}$ fixed as $d\rightarrow 0$.
The Green's function G then satisfies the Helmholtz equation:
$$
(\kappa^{2}-\bigtriangledown_{x})\bar{G}({\bf x},{\bf x'})=\delta^{D}
({\bf x}-{\bf x'}),\eqno(2.7)
$$
which is a well known result. Here,$\bigtriangledown=-\sum_{a}p_{a}^{2}
\equiv-{\bf p}^{2}.$

The operator formulation of the random walk problem does not easily
generalize to strings;instead, it is better to use the path integral
approach. It is convenient to introduce link vectors ${\bf e}_{a}$
of unit length in the positive ``a'' direction and vectors ${\bf e}_{a,j}
=j{\bf e}_{a}$ representing unit steps in $+$ or $-$ ``a'' direction,where
$j=\pm 1$. The correlation function G is then given by
$$
G=\sum_{m=0}^{\infty}G^{(m)},\eqno(2.8a)
$$
where
$$
G^{(m)}=\int\prod_{m'=1}^{m-1}d^{D}{\bf X}(m')\prod_{m''=1}^{m}d^{D}
{\bf P}(m'')exp(iI^{(m)}-(m-1)\alpha d),\eqno(2.8b)
$$
and,
$$
exp(iI^{(m)})=\sum_{\bf e}exp\Bigl(i\sum_{m'=1}^{m}{\bf P}(m').
({\bf X}(m')-{\bf X}(m'-1)-d{\bf e}_{a,j}(m')\Bigr).\eqno(2.8c)
$$
In this equation,integration over ${\bf P}$ enforces the constraint
that adjacent ${\bf X}'s$ on the lattice differ by one of the link
vectors $d{\bf e}_{a,j}$.It is also understood that the initial and
final points of the path,${\bf X}(0)$ and ${\bf X}(m)$, are fixed at
${\bf X}$ and ${\bf X'}$.

In the continuum limit, the discrete parameter m is replaced by the
continuous parameter $\tau=md$. Again expanding to second order in d
according to
$$
\sum_{aj}exp(-id{\bf e}_{aj}.{\bf P})\approx exp\Bigl(ln(2D)-\frac
{d^{2}}{2D}{\bf P}^{2}\Bigr),\eqno(2.9a)
$$
we have the continuum action
$$
iI=\int d\tau(2i{\bf X'}.{\bf P}-{\bf P}^{2}-\kappa^{2}),
\eqno(2.9b)
$$
which descibes a simple Gaussian model, in agreement with eq.(2.7).
Here, $\kappa^{2}$ is a constant.

Next, we consider a non-backtracking random walk on the lattice. This
is a Markov process, and it can be handled by slightly generalizing the
operator approach we have used for the unrestricted random walk. Now
we pick a point ${\bf X}$ on the path,with the previous step leading
to it given by ${\bf e}_{a,j}$ and the following step by ${\bf e}_
{b,k}$ (Fig.3). If the path is non-backtracking, $a=b,k=-j$ is not
allowed. To take this into account, the state representing ${\bf X}$
should have the memory of the previous step. This is accomplished by
attaching the indices ``a'' and ``j'' of the previous step to the
state $|{\bf X}>$ by defining
$$
|{\bf X},a,j>\equiv |{\bf X}>\otimes |a,j>,\eqno(2.10)
$$
where $|a,j>$ are orthonormal states. The two point correlation function
(see eq.(2.5)) is then given by
$$
G({\bf X},a,j,{\bf X'},b,k)=<{\bf X'},b,k|(1-e^{-\alpha d}T)^{-1}
|{\bf X},a,j>,\eqno(2.11a)
$$
where the 2D by 2D matrix T is given by
$$
T_{bk,aj}=exp(idp_{b}k)(1-\delta_{a,b}\delta_{k,-j}),\eqno(2.11b)
$$
and $p_{a}=-i\frac{\partial}{\partial x_{a}}$ as before. It is easy to
check that T does exactly what it is supposed to do; it moves ${\bf X}
$ by one step in all directions except the backtracking one.

To gain some insight into what is going on, it is useful to diagonalize
T for fixed ${\bf p}$.Defining $\omega=2\sum_{a}cos(dp_{a})$, the four
distinct eigenvalues are given by
$$
\eqalignno{
\lambda^{(1)}=&\frac{\omega}{2}+\frac{1}{2}(\omega^{2}-8D+4)^{1/2},\cr
\lambda^{(2)}=&\frac{\omega}{2}-\frac{1}{2}(\omega^{2}-8D+4)^{1/2},\cr
\lambda^{(3)}=&+1,\cr
\lambda^{(4)}=&-1.&(2.12)\cr}
$$
The corresponding eigenvectors can be written in the following form:
$$
|i>=\sum_{aj}\alpha_{a,j}^{(i)}|a,j>,\eqno(2.13a)
$$
where the $\alpha$'s are constants, and the index $i=1,2,3,4$ labels
the eigenvectors corresponding to the eigenvalues $\lambda^{(1,2,3,
4)}$. The equations satisfied by the $\alpha$'s are
$$
\eqalignno{
\alpha_{a,j}^{(1,2)}=& 1-\lambda^{(1,2)}exp(idjp_{a}),\cr
\sum_{aj}\alpha_{a,j}^{(3,4)}&= 0,\cr
\alpha_{a,-1}^{(3)}=& -exp(-idp_{a})\alpha_{a,1}^{(3)},\cr
\alpha_{a,-1}^{(4)}=& exp(-idp_{a})\alpha_{a,1}^{(4)}.&(2.13b)\cr}
$$
 Since T is not hermitian,the eigenvectors are not
orthogonal. This apparent lack of symmetry between the past and the
 future is due to the way we chose to label the states, and it does not
have any physical consequence. For example,physical quantities like
the eigenvalues are real. We also note that the eigenvectors
 corresponding to $i=1,2$ are non-degenerate, whereas the
 eigenvectors corresponding to $i=3,4$ are each $D-1$ fold
degenerate. It is  useful to expand $\lambda^{(1,2)}$ to second
order in d in anticipation of the continuum limit:
$$
\eqalignno{
\lambda^{(1)}\cong & 2D-1-\frac{2D-1}{2D-2}d^{2}{\bf p^{2}},\cr
\lambda^{(2)}\cong & 1+\frac{1}{2D-2}d^{2}{\bf p^{2}}.&(2.14)\cr}
$$
In the limit $d\rightarrow 0$,the eigenvector corresponding to the
eigenvalue $\lambda^{(1)}$ dominates over the rest; and therefore, in the
continuum limit, we could drop all the eigenvectors except the first
one.However, we shall see later that this is no longer true in the case
of strings; accordingly, we postpone the discussion of the continuum
limit to the next section. In passing, we note that
the zeroth order term in the expansion of $\lambda^{(1)}$ is reduced
 from 2D to $2D-1$  due to the fact that the step in the
backtracking direction  is no longer allowed.

 Let us now recast eq.(2.11) into the path integral language, with the
view of generalizing it to strings later. In the unconstrained random walk
problem, only ${\bf X}$ and ${\bf P}$ were needed as variables of
integration.However, since we now have discrete indices a and j to deal
with, additional variables of integration have to be introduced. These
 variables could be either bosonic or fermionic, and in one and two
dimensional spaces, they are known to be equivalent.(Here we are referring
to the dimensions of the world line and the world sheet). We find that
fermionic variables have several advantages over the bosonic ones;
for one thing, the formulas look much simpler, and also finitely indexed
fermions naturally span a finite dimensional space due to the exclusion
principle. Let us therefore introduce fermionic c-number variables
$\bar{\eta}_{a,j}(m)$ and $\eta_{a,j}(m)$ to keep track of the indices
a and j. Then the $I^{(m)}$ of eq.(2.8b) can in the non-backtracking case
 be written as
$$
\eqalignno{
iI^{(m)}&= \sum_{m'=1}^{m-1}\Bigl(i{\bf P}(m').({\bf X}(m')-{\bf X}
(m'-1))
\cr &+\frac{i}{2}\sum_{aj}(\bar{\eta}_{a,j}(m')\eta_{a,j}(m'+1)-
\bar{\eta}_{a,j}(m'+1)\eta_{a,j}(m'))\cr &+\sum_{aj,bk}\bar{\eta}_
{b,k}(m')M_{bk,aj}(m')\eta_{a,j}(m')\Bigr).&(2.15)\cr}
$$
To establish connection with the operator formulation of the same
problem and to find M, the transfer matrix for the $\eta's$ should be
 identified with
the matrix T of (2.11b). Remembering that $T_{bk,aj}(m)$ maps the state
\, $|a,j>$ at step m into the state$|b,k>$ at step m+1, we must have
$$
\eta(m+1)=T(m)\eta(m),\eqno(2.16a)
$$
where we have used the matrix notation. Comparing this with the
equations of motion for $\eta$ that follow from the lagrangian of
eq.(2.15),
$$
\frac{i}{2}(\eta(m+1)-\eta(m-1))+M(m)\eta(m)=0,\eqno(2.16b)
$$
gives us the relation between M and T:
$$
\frac{i}{2}(T(m)-T^{-1}(m-1))+M(m)=0.\eqno(2.16c)
$$
Parenthetically, the expression for T given in (2.11b) should be
slightly modified: The operator $p_{b}$ should be replaced by the
integration variable $P_{b}(m)$. This is why T acquires an apparent
m dependence.

The action of eq.(2.15),with M given by (2.16c), is still not
completely equivalent to the operator approach; it has to be
supplemented by an important boundary condition. The transfer matrix
T(m) maps a single fermion at step m, represented by $\eta(m)$,into
again a single fermion at step $m+1$. On the other hand,once we have
a functional integral, the resulting field theory will have all the
 states containing arbitrary number of fermions, up to the maximum
 allowed by fermi statistics. These extra states are unwanted; for
 example,in the zero fermion sector(vacuum), the T matrix is trivial
(unity),and this sector therefore describes the unconstrained random
 walk.To eliminate these
unwanted sectors, we note that the fermion number is conserved, and
sectors with different fermion numbers do not mix. We can therefore
impose the boundary condition that the initial state has fermion
number one; this condition will be preserved by the dynamics. This
initial condition is the extra ingredient that supplements the action
given by eq.(2.15).
\vskip 9pt
\noindent {\bf 3.Non-Backtracking Strings}
\vskip 9pt

In this section, we will try to extend the results obtained so far to
non-backtracking strings. We have the same D dimensional lattice
of the last section,and as a preliminary exercise, we first start with
the ordinary unconstrained sum over surfaces traced by a string(string
world sheet). At this point, we have to define what we mean by a surface.
It is easy to think of several reasonable definitions; most of these,
possibly even all of them, will have the same continuum limit.Here we
choose a  simple definition: Consider the coordinate variable
$X_{a}(m,n)$, parametrized by two integers m and n. The surface is
defined by requiring that two adjacent points on it are connected by
a link vector, hence the following equations must hold for some $a,j,
a',j'$:
$$
\eqalignno{
{\bf X}(m,n)-{\bf X}(m-1,n)&=d{\bf e}_{a,j}^{(1)}(m,n),\cr
{\bf X}(m,n)-{\bf X}(m,n-1)&=d{\bf e}_{a',j'}^{(2)}(m,n)
,&(3.1)\cr}
$$
where the ${\bf e}'s$ are the same unit vectors defined in the last
section, and superscripts (1) and (2) are attached to distinguish
between them. These constraints can be implemented by introducing
the lagrange multipliers ${\bf P}^{(1)}$ and ${\bf P}^{(2)}$, and the
partition function Z as a sum over surfaces is then given by
$$
\eqalignno{
Z=&\int\prod_{mn}d^{D}{\bf X}(m,n)d^{D}{\bf P}^{(1)}(m,n)d^{D}
{\bf P}^{(2)}(m,n)\cr &\sum_{aja'j'}exp\Bigl(i\sum_{mn}
({\bf P}^{(1)}
(m,n).({\bf X}(m,n)-{\bf X}(m-1,n)-d{\bf e}_{a,j}^{(1)}(m,n))\cr &
+{\bf P}^{(2)}(m,n).({\bf X}(m,n)-{\bf X}(m,n-1)-d{\bf e}_{a',j'}
^{(2)}(m,n))\Bigr).&(3.2)\cr}
$$
Here, for simplicity, we have not introduced a term propotional to
the area (Nambu term) in the exponential, since, as we shall see,
such a term is automatically generated in any case.
To arrive at a correlation function anologous to G of the random
walk problem(see eq.(2.8)), again one has to fix the initial and the
final ${\bf X}'s$ and sum over the number of steps it takes to get
from one to the other. Let us now consider the naive(formal) continuum
limit of this equation. The action in this limit is gotten by expanding
 in powers of d, and by dropping terms that vanish in the limit
$d\rightarrow 0$. It is of course necessary to scale the field
 variables correctly to end up with the correct continuum kinetic
energy terms. Defining the coordinates on the world sheet by
$$
\sigma_{1}=md, \sigma_{2}=nd,\eqno(3.3a)
$$
and scaling by
$$
\eqalignno{
{\bf P}^{(1,2)}&\rightarrow (2D)^{1/2}{\bf P}^{(1,2)},\cr
{\bf X}&\rightarrow (2/D)^{1/2}d  \ \ {\bf X},&(3.3b)\cr}
$$
we arrive at the result,
$$
iI\rightarrow \int d^{2}\sigma\sum_{l=1,2}(2i{\bf P}^{(l)}
(\sigma).\frac{\partial{\bf X}(\sigma)}{\partial\sigma_{l}}
-({\bf P}^{(l)}(\sigma))^{2}).\eqno(3.4)
$$
A few comments on this result are in order:\\
a)Integrating over ${\bf P}$ reproduces the action for the standard
bosonic string, gauge fixed in the conformal gauge.However,the gauge
fixing terms(ghosts) are missing. This  presumably comes about
because the naive continuum limit misses subtle quantum effects.
It is easily fixed by the standard procedure of introducing the world
sheet metric, casting (3.4) into reparametrization invariant form, and
finally fixing the gauge(14).\\
b)The continuum result is rotationally invariant on the world sheet.
This is of course essential for being able to generalize it to a
reparametrization invariant form; since we have a constant background
metric on the world sheet, the only remnant of reparametrization
 invariance is a global rotation invariance, and this should remain
unbroken. There is even a discrete residue of this symmetry in the
case of the lattice: Eq.(3.2) is invariant under
$m\leftrightarrow n, 1\leftrightarrow 2$. When we introduce
fermionic variables later,this symmetry will generalize to a symmetry
between the right and left chiral sectors of the model.\\
c)It was pointed out earlier that the definition of a world sheet surface
was non-unique. In fact,the concept of  surface is not central in our
approach. All we require is that, upon identifying n with a discrete time
variable,the coordinate vector ${\bf X}(m,n)$ should trace all the
acceptable Wilson lines(loops) for a fixed n and variable m. On account
 of the discrete symmetry discussed above,this should be true also for
fixed m and variable n. Therefore, in our approach, the world sheet can
be thought of as a direct product of two sets of  Wilson
 lines, traced by ${\bf X}(m,n)$ for either fixed m and variable n,
 or for fixed n and variable m respectively.

 As it stands, eq.(3.2) does not generate a
satisfactory string theory, since backtracking Wilson loops
 are present in the sum. To eliminate them, we proceed as in the random
walk problem discussed in section 2. To the first set of Wilson lines,
we attach the fermionic variables $\eta_{a,j}^{(1)}(m,n)$, and to the
 second set the variables $\eta_{a,j}^{(2)}(m,n)$; these variables
``guide'' the path so that it is non-backtracking. For a single path,
the transfer matrix T for the $\eta's$  is given by eq.
(2.11b). In the present case, we have two transfer matrices $T^{(1,2)}$
, each given by equation (2.11b), with $p_{b}$ replaced by $P_{b}^{(1)}
(m,n)$ and $P_{b}^{(2)}(m,n)$ respectively. The partition function is
therefore given by
$$
\eqalignno{
Z=&\int{\cal D}X{\cal D}P^{(1)}{\cal D}P^{(2)}{\cal D}\eta^{(1)}
{\cal D}\eta^{(2)}{\cal D}\bar{\eta}^{(1)}
{\cal D}\bar{\eta}^{(2)} exp(iI),\cr
&where \cr I=&I_{1}+I_{2}+I_{3},\cr& and,\cr
I_{1}=&\sum_{mn}\Bigl({\bf P}^{(1)}(m,n).({\bf X}(m,n)-{\bf X}(m-1,n))
+{\bf P}^{(2)}(mn).({\bf X}(m,n)-{\bf X}(m,n-1))\Bigr),\cr
I_{2}=&\frac{i}{2}\sum_{mn}\Bigl(\bar{\eta}^{(1)}(m,n)\eta^{(1)}
(m+1,n)-\bar{\eta}^{(1)}(m+1,n)\eta^{(1)}(m,n)\cr &+\bar{\eta}^{(2)}
(m,n)\eta^{(2)}(m,n+1)-\bar{\eta}^{(2)}(m,n+1)\eta^{(2)}(m,n)\Bigr),
\cr
I_{3}=&\sum_{mn}\sum_{l=1,2}\bar{\eta}^{(l)}(m,n)M^{(l)}(m,n)\eta
^{(l)}(m,n).&(3.5)\cr}
$$
In this equation, we have again adopted the matrix notation and
suppressed the indices a and j attached to the $\eta's$.From the
definition of T and the equation of motion for $\eta$, the following
relation between T and M follows(see eq.(2.16b)):
$$
\eqalignno{
\frac{i}{2}\Bigl(T^{(1)}(m,n)-(T^{(1)}(m-1,n))^{-1}\Bigr)
+M^{(1)}(m,n)&=0,\cr
\frac{i}{2}\Bigl(T^{(2)}(m,n)-(T^{(2)}(m,n-1))^{-1}\Bigr)
+M^{2}(m,n)&=0,&(3.6a)\cr}
$$
where T is given by
$$
T^{(1,2)}_{bk,aj}(m,n)=exp(idkP_{b}^{(1,2)}(m,n))(1-\delta_{a,b}
\delta_{k,-j}).\eqno(3.6b)
$$

Eqn's (3.5) and (3.6) describe the dynamics of non-bactracking paths
in the functional approach. As explained at the end of section 2, they
have to be supplemented by boundary conditions on the fermion number.
These boundary conditions must ensure that one and only one fermion is
attached to each of the fixed m and  the fixed n trajectories.
In the case of random walk, the existence of a conserved fermion
number was crucial in preserving the boundary condition along the
trajectory. Here, in contrast, we have two sets of conserved fermion
number \underline{densities} $N^{(1)}(m,n)$ and $N^{(2)}(m,n)$, where
the first denotes the number of fermions propagating along the link
connecting (m,n) to (m+1,n), and the second those propagating along
the link (m,n) to (m,n+1). The conservation laws, which follow from
(3.5), are
$$
N^{(1)}(m+1,n)=N^{(1)}(m,n),\: N^{(2)}(m,n+1)=N^{(2)}(m,n).\eqno(3.7)
$$
 The following boundary conditions are then imposed: For a
fixed initial value $m_{i}$ of m, we require that
$$
N^{(1)}(m_{i},n)=1,\eqno(3.8a)
$$
and, for a fixed initial value $n_{i}$ of n, the condition is
$$
N^{(2)}(m,n_{i})=1.\eqno(3.8b)
$$
These conditions are preserved along each trajectory by virtue of the
 conservation laws, and they eliminate unwanted sectors, including the
 sector $N^{(1)}=N^{(2)}=0$, which corresponds to the standard bosonic
 string of eq.(3.2).

Is the string theory defined by eq.(3.5) the QCD string? In general, we
cannot answer this question for a string that lives on a lattice. However,
we expect to show in a future publication(10) that in the limit of
continuous time, still keeping space latticized, the string theory we
have constructed is equivalent to a gauge theory on a space lattice. This
theory differs from the standard hamiltonian lattice QCD in having
 non-minimal interaction terms, although the formal continuum limit is
 the same. It is therefore reasonable to hope that the continuous space-time
 limit of our model, which
we will study in the next section, is in fact the large N limit of QCD.
\vskip 9pt
\noindent {\bf 4. The Continuum Limit}
\vskip 9pt

In this section, we shall study the naive(or formal) continuum limit of
eq.(3.5). The world sheet coordinates $\sigma_{1,2}$ are defined and
bosonic  variables ${\bf X}$ and ${\bf P}$ are scaled as before
(eqs.(3.3a,b)), whereas the fermionic variables are scaled by
$$
\eta^{(1,2)}(m,n)\rightarrow d^{1/2}\eta^{1,2}(\sigma_{1},\sigma_{2}).
\eqno(4.1)
$$
The continuum limits of $I_{1,2}$ are straightforward to compute:
$$
\eqalignno{
I_{1}&\rightarrow \int d^{2}\sigma\sum_{l=1,2}{\bf P}^{(l)}.\frac{
\partial{\bf X}}{\partial\sigma_{l}},\cr
I_{2}&\rightarrow i\int d^{2}\sigma\sum_{l=1,2}\bar{\eta}^{(l)}\frac{
\partial\eta^{(l)}}{\partial\sigma_{l}}.&(4.2)\cr}
$$
Finding the limit of $I_{3}$ involves little a bit more work. It is
convenient to first diagonalize T (see eqs.(2.12) and (2.13)) before
taking the $d\rightarrow 0$ limit, and then compute M through eq.(2.16c).
In this calculation, only terms up to first order in d are needed, since
higher order terms do not contribute in this limit. To first order, the
eigenvalues $\Lambda_{i}$ of M are
$$
\eqalignno{
\Lambda_{1}&\cong\frac{i}{2}(\frac{1}{2D-1}-2D+1),\cr
\Lambda_{2}&\cong 0,\cr
\Lambda_{3}&=0,\cr
\Lambda_{4}&=0,&(4.3a)\cr}
$$
and the right and the left eigenvectors $|i>$ and $<i|$ are given as a
sum over states $|a,j>$ and $<a,j|$ (see eq.(2.10) for the definition):
$$
\eqalignno{
|1>&\cong (2D)^{-1/2} \sum_{aj}(1+id\;\frac{2D-1}{2D-2}P_{a}j)|a,j>,\cr
<1|&\cong (2D)^{-1/2} \sum_{aj}(1+\frac{id}{2D-2}P_{a}j)<a,j|.&(4.3b)
\cr}
$$
The states $|2>, |3>$ and $|4>$ are not needed, since, the corresponding
eigenvalues being 0, they do not appear in the expansion of M. Notice that
as T and M are not hermitian, the left eigenvector $<1|$ is not the
conjugate of the right eigenvector $|1>$. Also, the terms T and
$T^{-1}$ in (3.6a) have different arguments, and this causes some confusion
 about what the argument of $P_{a}$ should be in eq.(4.3b). However,
replacing $m-1$ by m, or $n-1$ by n changes the final result only by
terms of order $d^{2}$ and therefore it is permissible. The matrix M, again
to first order in d, is given by
$$
M^{(1,2)}_{bk,aj}=\frac{i\Lambda_{1}}{2D}(1+id\;\frac{2D-1}{2D-2}P^{(1,2)}
_{b}k+\frac{id}{2D-2}P^{(1,2)}_{a}j).\eqno(4.4)
$$
Substituting eq.(4.4) in (3.5) then gives the continuum limit of $I_{3}$.
 This term can be simplified substantially by defining a new set of
$\eta's$ by
$$
\eqalignno{
s^{(l)}&=(2D)^{-1/2}\sum_{aj}\eta_{a,j}^{(l)},\cr
v_{a}^{(l)}&=2^{-1/2}\sum_{j}j\eta_{a,j}^{(l)},\cr
w_{c}^{(l)}&=\sum_{aj}f_{c,a}\eta_{a,j}^{(l)},&(4.5a)\cr}
$$
and a new set of $\bar{\eta}'s$ through the same equations.
Here, $l=1,2$ as before, and the index c takes on $2D-2$
values.The real constants $f_{c,a}$ satisfy the conditions
$$
\eqalignno{
\sum_{a}f_{c,a}f_{c',a}&=\delta_{c,c'},\cr
\sum_{a}f_{c,a}&=0.&(4.5b)\cr}
$$
Eqn.'s (4.5a,b) are a set of orthogonal transformations to a new set of
variables $s^{(l)},v_{a}^{(l)}$ and $w_{c}^{(l)}$, which diagonalize M,
and which give the following simple result for $I_{3}$:
$$
I_{3}\rightarrow\Lambda_{1}\int d^{2}\sigma\sum_{l}\Bigl(\frac{i}{d}
\bar{s}^{(l)}s^{(l)}-\frac{1}{D^{1/2}(2D-2)}\sum_{a}(\bar{s}^{(l)}
v_{a}^{(l)}+(2D-1)\bar{v}_{a}^{(l)}s^{(l)})P_{a}^{(l)}\Bigr).\eqno(4.6)
$$
Now that we have the continuum limit, let us discuss some of its
features:\\
a)The theory is rotation invariant in the D dimensional target space,
provided that we assign the following transformation laws to the fields:
$P_{a}^{(1,2)},v_{a}^{(1,2)}$ and $\bar{v}_{a}^{(1,2)}$ are D vectors,
and the rest, $s^{1,2}$ and $w_{c}^{(1,2)}$, are all scalars.This
 assignment
follows uniquely from the fact that $X_{a}$ is a D vector. If so desired,
 one can then introduce the Minkowsky metric in the target space
and end up with a Lorentz invariant theory.\\
b)The fields $w_{c}^{(1,2)}$ decouple from the rest of the theory and
become free fields. Of course, when a metric is introduced on the world
sheet to restore reparametrization invariance, they will couple to this
metric and will contribute to the value of the central charge and to the
calculation of higher order terms in $1/N$. Since we will not study these
more subtle issues in this paper, from now on these fields will be
 dropped.\\
c)Since the background metric is flat, depending on its signature,we expect
 to have either Lorentz or rotation symmetry on the world sheet. To see how
this works, we first consider the kinetic energy terms for the field
$\eta$ in eq.(4.2). From the structure of these terms, it is clear that
$\eta^{(1)}$ and $\eta^{(2)}$ have to be identified with the the two chiral
(left and right) components of a world sheet spinor. Accordingly, we define
 the following two component spinors:
$$
\eqalignno{
\eta_{a,j}&\equiv \left(
       \matrix{\eta_{a,j}^{(1)}\cr
               \eta_{a,j}^{(2)} \cr}\right),\cr
\bar{\eta}_{a,j}&\equiv (\bar{\eta}_{a,j}^{(2)},\bar{\eta}_{a,j}^{(1)}),
&(4.7a)\cr}
$$
with a similar definition for s and $v_{a}$. $I_{2}$ then becomes the
standard Dirac kinetic energy term:
$$
I_{2}=i\int d^{2}\tau\sum_{aj}\bar{\eta}_{a,j}\gamma^{\mu}\frac{\partial
\eta_{a,j}}{\partial\tau^{\mu}},\eqno(4.7b)
$$
where $\mu=0,1$, and,
$$
\eqalignno{
\gamma^{0}&=\sigma_{x},\; \gamma^{1}=i\sigma_{y},\cr
\tau^{0}&=\sigma_{1}+\sigma_{2},\; \tau^{1}=\sigma_{1}-\sigma_{2}.&(4.7c)
\cr}
$$
In this equation, $\sigma_{x,y}$ are Pauli matrices, not to be confused
with $\sigma_{1,2}$, the world sheet coordinates. From the structure
of the Dirac equation, it is clear that $\sigma_{1,2}$ are the lightcone
coordinates, in contrast to $\tau^{0,1}$, the Cartesian coordinates. It
then follows that the world sheet metric is Minkowskian, which we have
already anticipated by writing the path integral (3.5) in the Feynman
form, with a factor ``i'' in front of the action I in the exponential.
We also need the covariant two-vecor components $P_{a}^{\mu}$ of momentum
 in terms of the lightcone components $P_{a}^{1,2}$:
$$
P_{a}^{0}=P_{a}^{(1)}+P_{a}^{(2)},\;P_{a}^{1}=P_{a}^{(1)}-P_{a}^{(2)}.
\eqno(4.7d)
$$
With these assignments of transformation laws for the fields, the second
 and the third terms on the right hand side of eq.(4.6) can be written in
world sheet Lorentz invariant form:
$$
\eqalignno{
\sum_{al}\bar{v}_{a}^{(l)}s^{(l)}P_{a}^{(l)}&=\frac{1}{2}\sum_{a,\mu}
\bar{v}_{a}\gamma^{\mu}sP_{a,\mu},\cr
\sum_{al}\bar{s}^{(l)}v_{a}^{(l)}P_{a}^{(l)}&=\frac{1}{2}\sum_{a,\mu}
\bar{s}\gamma^{\mu}v_{a}P_{a,\mu}.&(4.8)\cr}
$$
However, the first term on the right hand side of (4.6), proportional
to,
$$
S\equiv\frac{1}{d}\sum_{l}\bar{s}^{(l)}s^{(l)},\eqno(4.9)
$$
is not world sheet Lorentz invariant as it stands. We will return
 to this problem later.\\
d)All of the terms but one in the action have dimensionless coupling
 constants and therefore, they are classically scale invariant.
The only exception is again S, the Lorentz non-invariant term discussed
 above.\\
e)The action is parity invariant (left-right symmetric) on the world sheet.

We will now show that the troublesome term S can be cast into a form
that is both scale and Lorentz invariant. To be able to do this, it is
important to realize that the form of the action is not unique;
one can  multiply any one of its terms by one of the
number density operators constrained to be unity(eqs.(3.8a,b)).
Expressed in terms of the fermionic variables on the lattice, these
constraints can be written as
$$
\eqalignno{
1=&N^{(l)}(m,n)\rightarrow\sum_{aj}\bar{\eta}_{a,j}^{(l)}(m,n)\eta_{aj}
^{(l)}(m,n)\cr
=&\bar{s}^{(l)}s^{(l)}+\sum_{a}\bar{v}_{a}^{(l)}v_{a}^{(l)}+\sum_{c}
\bar{w}_{c}^{(l)}w_{c}^{(l)}.&(4.10a)\cr}
$$
In the continuum limit, this equation becomes
$$
d(\bar{s}^{(l)}(\tau)s^{(l)}(\tau)+\sum_{a}\bar{v}_{a}^{(l)}(\tau)
v^{(l)}(\tau))\cong 1,\eqno(4.10b)
$$
where, as explained earlier, the term involving w's was dropped. The
 factor of d comes from the scaling of the fields in passing to the
 continuum limit. The term S, which is quadratic in field variables,
can now be converted into a quartic term by multiplying it by a
suitable combination of terms that appear on the right hand side of
(4.10b):
$$
\eqalignno{
S&=\frac{1}{d}(N^{(1)}\bar{s}^{(2)}s^{(2)}+N^{(2)}\bar{s}^{(1)}
s^{(1)})\cr &\rightarrow \frac{1}{2}\bar{s}\gamma^{\mu}s\;
(\bar{s}\gamma_{\mu}s+\sum_{a}\bar{v}_{a}\gamma_{\mu}v_{a}).&(4.11)
\cr}
$$
Finally, putting together eqs.(4.7b,d),(4.8) and (4.11), we arrive
at the following expression for the continuum action:
$$
\eqalignno{
I&=\int d^{2}\tau \Bigl(P_{a}^{\mu}\;\partial_{\mu}X_{a}
+c_{1}\;P_{a}^{\mu}\;P_{a,\mu}+i\bar{s}\;\gamma^{\mu}\partial_{\mu}s
+i\bar{v}_{a}\;\gamma^{\mu}\partial_{\mu}v_{a}\cr &+c_{2}\;\bar{s}
\gamma^{\mu}s\;\bar{s}\gamma_{\mu}s+c_{3}\;\bar{s}\gamma^{\mu}s\;
\bar{v}_{a}\gamma_{\mu}v_{a}+c_{4}\;\bar{v}_{a}\gamma^{\mu}v_{a}\;
\bar{v}_{b}\gamma_{\mu}v_{b}\cr &+c_{5}\;\bar{s}\gamma^{\mu}v_{a}\;
P_{a,\mu}+c_{6}\;\bar{v}_{a}\gamma^{\mu}s\;P_{a,\mu}+c_{7}\;\bar{s}
\gamma^{\mu}v_{a}\;\bar{v}_{a}\gamma_{\mu}s\Bigr),&(4.12a)\cr}
$$
where the summation convention for repeated indices is used, and the
 partials are with respect to the variable $\tau$. The
constants ``c'' are given by
$$
\eqalignno{
c_{1}&=c_{4}=c_{7}=0,\;c_{2}=c_{3}=\frac{1}{2}i\Lambda_{1},\cr
c_{5}&=-\frac{\Lambda_{1}}{2D^{1/2}(2D-2)},\;c_{6}=-\frac{\Lambda
_{1}(2D-1)}{2D^{1/2}(2D-2)}.&(4.12b)\cr}
$$

We have written eq.(4.12a) deliberately with arbitrary constants c;
written in this fashion, it is the most general local action
invariant under the symmetries listed after eq.(4.6). In the naive
continuum limit, these constants have the values given in eq.(4.12b).
However, the naive continuum limit is clearly not the final answer,
since under the flow of the renormalization group, these constants
will be driven from their initial values given by (4.12b) to some
 fixed point corresponding to different values for these constants.
 If this fixed point represents
pure QCD, it is natural to expect it to have the following properties:\\
a)It should be conformally invariant.
Otherwise, renormalization would introduce a new scale in addition to
the slope parameter which multiplies the action and which we have set
equal to one. Since we know that pure QCD has only a single scale
parameter, conformal invariance has to be an essential feature of the
world sheet field theory.\\
b) The constant
$c_{1}$ and at least one of the constants $c_{5}$ or $c_{6}$ should be
different from zero at the fixed point. The non-vanishing of $c_{1}$ is
 necessary to have asymptotically linear Regge trajectories. With $c_{1}$
not zero, we can integrate over P and recover the standard quadratic term
in X. On the other hand, if both
$c_{5}$ and $c_{6}$ vanished, the fermionic and the bosonic sectors
of (4.12a) would decouple from each other, and the bosonic sector by
itself would simply reproduce the standard string model with purely linear
trajectories.( Here the labels bosonic and fermionic refer to the world
sheet variables). Since we already know that the standard string model does
not describe QCD, it follows that either $c_{5}$or $c_{6}$ should be
 different from zero.  If this is indeed the case, the resulting
trajectories will be in general curved because of the coupling between
P and the bilinears in fermions. As explained in the introduction, one
expects QCD trajectories to be curved, and so this is a step in the right
direction.\\
c)The action given by (4.12) does not look hermitian.Some of this apparent
non-hermiticity can be eliminated by redefinitions of fields. On the other
hand, an imaginary value for $c_{2}$, as given by eq. (4.12b), is a genuine
 violation of hermiticity. However, this is only the initial value of
$c_{2}$ and not its value at the fixed point.
Since the model should be unitary, we expect $c_{2}$ to become real at the
fixed point. A related question of unitarity is whether the ghosts due to the
Minkowski metric in target space decouple. We have nothing to say about this
important problem in this paper.In any case,
it is clear that to  have any chance of a viable theory, the generalized
 Thirring model (4.12a) must have a fixed point with the properties listed
 above. To investigate the fixed points of this model is
therefore a task of great importance for the future.

We close this section with a discussion of the continuum limit of the
constraints (3.8), which supplement the action of (4.12). We remind the
 reader that these constraints are necessary to eliminate the spurious
sectors present in the model.They could also play an important role in
making the theory unitary by eliminating states with negative norm. In
order to write them in covariant form, it is convenient to introduce the
following vector current:
$$
J^{\mu}\equiv \bar{s}\gamma^{\mu}s+\sum_{a}\bar{v}_{a}\gamma^{\mu}v_{a},
\eqno(4.13a)
$$
whose light cone components turn out to be the fermion densities of
eq.(3.7):
$$
\eqalignno{
N^{(1)}&\rightarrow J^{(-)}=\frac{1}{2}(J^{0}-J^{1}),\cr
N^{(2)}&\rightarrow J^{(+)}=\frac{1}{2}(J^{0}+J^{1}).&(4.13b)\cr}
$$
The continuum analogues of eqns.(3.8a,b) are then imposed on $j^{(\mp)}$
at fixed initial values of $\sigma_{1}=\tau^{(+)}=\frac{1}{2}(\tau^{0}+
\tau^{1})$ and $\sigma_{2}=\tau^{(-)}=\frac{1}{2}(\tau^{0}-\tau^{1})$
respectively. For these conditions to be preserved along a trajectory,
J should be chirally conserved:
$$
\partial_{\mu}J^{\mu}=0,\;\varepsilon^{\mu\nu}\partial_{\mu}J_{\nu}
=0,\eqno(4.14a)
$$
or, in terms of the chiral components,
$$
\partial_{+}J^{(-)}=0,\;\partial_{-}J^{(+)}=0.\eqno(4.14b)
$$
These are then the continuum version of eqn.(3.7). The remaining problem
is how to impose the analogues of (3.8a,b) on J. In the quantized theory,
J is an operator, and it is well known that imposing  conditions such as
eq.(3.8) directly on the operators leads to inconsistencies. Instead, one
has to impose these conditions on the physically acceptable states,
using only the positive and zero modes of J. If $\tau^{(\pm)}$ are
compactified on an interval of length $2\pi$, we can define the Fourier
modes of J by
$$
J^{(\pm)}(\tau)=\sum_{n}exp(in\tau^{(\pm)})J_{(n)}^{(\pm)},\eqno(4.15)
$$
and write down the conditions that a physical state $|>$ must satisfy:
$$
J_{(n)}^{(+)}|>=0,\;
J_{(n)}^{(-)}|>=0,\eqno(4.16a)
$$
for $n\geq 1$ and,
$$
J_{(0)}^{(+)}|>=|>,\;
J_{(0)}^{(-)}|>=|>,\eqno(4.16b)
$$
for $n=0$.

The conservation laws given by eq.(4.14) are the consequence of the
invariance of the action (4.12a) under both vector and axial U(1)
transformations. Under vector U(1), the fields transform as
$$
\eqalignno{
s&\rightarrow exp(i\delta_{v})s,\;\bar{s}\rightarrow \bar{s}\;
exp(-i\delta_{V}),\cr
v_{a}&\rightarrow exp(i\delta_{v})v_{a},\;\bar{v}_{a}\rightarrow
\bar{v}_{a}\; exp(-i\delta_{V}),&(4.17a)\cr}
$$
and under the axial U(1), the transformation is
$$
\eqalignno{
s&\rightarrow exp(i\gamma_{3}\delta_{A})s,\;\bar{s}\rightarrow
\bar{s}\;exp(i\gamma_{3}\delta_{A}),\cr
v_{a}&\rightarrow exp(i\gamma_{3}\delta_{A})v_{A},\;\bar{v}_{a}
\rightarrow \bar{v}_{a}\; exp(i\gamma_{3}\delta_{A}),&(4.17b)\cr}
$$
where $\gamma_{3}=i\gamma^{0}\gamma^{1}$. $J^{\mu}$ and its dual are
then the conserved currents resulting from these symmetries. We should
therefore add these to the list of symmetries of the action listed
after eq.(4.6). However, these symmetries are not automatic, since
it is well known that axial transformations of the type given by
eq.(4.17b) are in general anomalous. Such an anomaly would destroy the
chiral conservation laws and would prevent us from imposing the
constraints given by (4.16) consistently. To see how this anomaly comes
about, we define a fermionic supermultiplet Z built out $D+1$ doublets
s and $v_{a}$:
$$
Z\equiv \left(\matrix{s\cr v_{a}\cr}\right),\eqno(4.18a)
$$
and using auxilliary fields $W_{i,\mu}$, we rewrite eq.(4.12a) as
$$
\eqalignno{
I&=\int d^{2}\tau \Bigl(P_{a}^{\mu}\;\partial_{\mu}X_{a}+c_{1}\;
P_{a}^{\mu}\;P_{a,\mu}+c_{5}\;\bar{s}\gamma^{\mu}v_{a}\;P_{a,\mu}
+c_{6}\;\bar{v}_{a}\gamma^{\mu}s\;P_{a,\mu}\cr &+i\bar{z}\;\gamma^{\mu}
\partial_{\mu}z+\sum_{i}(\bar{z}K_{i}\gamma^{\mu}z\;W_{i,\mu}+
W_{i}^{\mu}\;W_{i,\mu})\Bigr),&(4.18b)\cr}
$$
where $K_{i}$ are suitable $D+1\otimes D+1$ matrices so that upon
integrating over the W's, (4.12a) is reproduced. For  fixed W, there
is an axial U(1) anomaly proportional to the trace of the field
strength built out of W. The condition for the absence of this
anomaly is that
$$
tr(K_{i})=0\eqno(4.19a)
$$
for all`` i''.At the fixed point, this implies the following two
 equations between the coupling constants that appear in eq.(4.12a):
$$
c_{2}=D^{2}c_{4},\;c_{3}=-2Dc_{4}.\eqno(4.19b)
$$
Clearly, the argument that led to (4.19b) is only semiclassical and
may be modified by higher quantum corrections. We went through this
exercise merely to demonstrate that chiral U(1) invariance imposes
further conditions on the parameters of the model.
\vskip 9pt
\noindent {\bf 5.Conclusions}
\vskip 9pt

The main result of this paper is the introduction of a new string
model. This new model was derived by restricting the world sheet to
surfaces traced by non-backtracking paths on a lattice and then taking
the continuum limit. It was argued that such a restriction was the
necessary minimum condition for the string theory to represent a
non-abelian gauge theory in the large N limit. To implement this
restriction on the allowed paths, fermionic variables on the world
sheet were introduced, and it was shown that the world sheet action of
the resulting string theory is the generalized Thirring model of
 eq.(4.12a). This action has to be supplemented by the conditions on
admissible states given by eqs.(4.16a,b). It was also argued that the
coupling constants of the model should correspond to a fixed point,
and this fixed point should satisfy a number of conditions, such as
conformal invariance, in order to make connection with QCD.

Much remains to be done to find out whether the model is viable. First,
one has to establish the connection between the latticized version of
 the string model and QCD on the lattice(10), and then, to see whether
the continuum string makes sense, one has to find out whether the
generalized Thirring model(4.12a) has a fixed point with the required
properties. Hopefully, the work presented here will stimulate research
along these directions.
\vskip 9pt
\noindent {\bf Acknowledgement}
\vskip 9pt

I would like to thank Elias Kiritsis for a useful conversation.
\newpage
{\bf References}
\begin{enumerate}
\item 1. G.'t Hooft, Nucl.Phys. {\bf B72}(1974) 461.
\item 2. See, for example, A.M.Polyakov, Nucl.Phys. {\bf B268}
(1986) 406.
\item 3. J.Polchinski,``Strings and QCD?'', Univ.of Texas preprint,
(June 1992), Phys.Rev.Letters {\bf 68} (1992) 1267.
\item 4. For some very recent work on large N limit of QCD, see
V.A.Kazakov and A.A.Migdal, ``Induced QCD at large N'', Princeton
preprint PUPT-1322,(June 1992), Yu.Makeenko,``Large N reduction, Master
Field and Loop Equations in Kazakov-Migdal Model'', ITER-YM-6-92(August
1992), D.Gross,``Some Remarks about Induced QCD'',Princeton Preprint,
PUPT-1335(August 1992).
\item 5. G.'t Hooft, Nucl.Phys. {\bf B75} (1974) 461.
\item 6. T.Eguchi, Phys.Letters {\bf B87} (1679) 91, D.Foerster,Phys.
Letters {\bf B87} (1979) 87, D.Weingarten, Phys.Letters {\bf B87}
(1979) 79.
\item 7. Yu.Makeenko and A.A.Migdal, Phys.Letters {\bf B88} (1979) 135.
\item 8. V.A.Kazakov, Phys.Letters {\bf B128} (1983) 316, V.I.Kostov,
Phys.Letters {\bf B138} (1984) 191.
\item 9. S.Mandelstam, Phys.Rev. {\bf D19} (1979) 2391.
\item 10. K.Bardakci, work in progress.
\item 11. R.Dashen and Y.Frishman, Phys.Rev. {\bf D11} (1975) 2781.
\item 12. K.Bardakci, Nucl.Phys. {\bf B238} (1984) 621.
\item 13. For a review of Brownian motion and random walk problems,
see the article by E.W.Montroll and B.J.West in ``Studies in
Statistical Mechanics and Fluctuation Phenomena'', E.W.Montroll and
J.L.Lebowitz editors, North Holland, Amsterdam (1979).
\item 14. A.M.Polyakov, Phys.Letters {\bf B103} (1981) 207.
\end{enumerate}

\end{document}